\documentclass[aps,preprint,amsmath,amssymb,showpacs]{revtex4}

\usepackage{graphicx}
\begin{document}

\title{Unparticle Physics in the Moller Scattering}

\author{\.{I}nan\c{c} \c{S}ahin}
\email[]{isahin@science.ankara.edu.tr} \affiliation{Department of
Physics, Faculty of Sciences, Ankara University, 06100 Tandogan,
Ankara, Turkey}

\author{Banu \c{S}ahin}
\email[]{dilec@science.ankara.edu.tr} \affiliation{Department of
Physics, Faculty of Sciences, Ankara University, 06100 Tandogan,
Ankara, Turkey}

\begin{abstract}
We investigate the virtual effects of vector unparticles in the
Moller scattering. We derive the analytic expression for scattering
amplitudes with unpolarized beams. We obtain $95\%$ confidence level
limits on the unparticle couplings $\lambda_{V}$ and $\lambda_{A}$
with integrated luminosity of $L_{int}=50, \, 500 \, fb^{-1}$ and
$\sqrt{s}=100, \, 300$ and $500$ GeV energies. We show that limits
on $\lambda_{V}$ are more sensitive than $\lambda_{A}$.

\end{abstract}


\maketitle

\section{Introduction}

In his recent papers Georgi \cite{Georgi, Georgi2} has proposed a
new scenario. In his proposal new physics contains both Standard
Model (SM) fields and a scale invariant sector described by
Banks-Zaks (BZ) fields \cite{BZ}. The two sectors interact via the
exchange of particles with a mass scale $M_{U}$. Below this large
mass scale interactions between SM fields and BZ fields are
described by non-renormalizable couplings suppressed by powers of
$M_{U}$ \cite{Georgi, Cheung}:

\begin{eqnarray}
\frac{1}{M_{U}^{d_{SM}+d_{BZ}-4}}O_{SM}O_{BZ}
\end{eqnarray}

The renormalization effects in the scale invariant BZ sector then
produce dimensional transmutation at an energy scale $\Lambda_{U}$
\cite{Weinberg}. In the effective theory below the scale
$\Lambda_{U}$, the BZ operators are embedded as unparticle
operators. The operator (1) is now matched to the following form,

\begin{eqnarray}
C_{O_{U}}\frac{\Lambda_{U}^{d_{BZ}-d_{U}}}{M_{U}^{d_{SM}+d_{BZ}-4}}O_{SM}O_{U}
\end{eqnarray}
here, $d_{U}$ is the scale dimension of the unparticle operator
$O_{U}$ and the constant $C_{O_{U}}$ is a coefficient function.

If unparticles exist, their phenomenological implications should be
discussed. In the literature, there have been many discussions which
investigate various features of unparticle physics \cite{Luo}. In
the some of these researches several unparticle production processes
have been studied. Possible evidence for this scale invariant sector
might be the signature of a missing energy. It can be tested
experimentally by examining missing energy distributions.

Other evidence for unparticles can be explored by studying its
virtual effects. Imposing scale invariance, the spin-1 unparticle
propagator is given by \cite{Georgi2,Cheung2}:

\begin{eqnarray}
\Delta(P^{2})^{\mu\nu}=i\frac{A_{d_{U}}}{2sin(d_{U}\pi)}(-P^{2})^{d_{U}-2}
\left(-g^{\mu\nu}+\frac{P^{\mu}P^{\nu}}{P^{2}} \right)
\end{eqnarray}
where,

\begin{eqnarray}
A_{d_{U}}=\frac{16\pi^{\frac{5}{2}}}{(2\pi)^{2d_{U}}}\frac{\Gamma(d_{U}+\frac{1}{2})}
{\Gamma(d_{U}-1)\Gamma(2d_{U})}
\end{eqnarray}

In this work we investigate virtual unparticle effects through
Moller scattering. We consider the following effective interaction
terms, first proposed by Georgi\cite{Georgi2}:

\begin{eqnarray}
i\frac{\lambda_{V}}{\Lambda_{U}^{d_{U}-1}}\bar{f}\gamma_{\mu}f
O_{U}^{\mu}+i\frac{\lambda_{A}}{\Lambda_{U}^{d_{U}-1}}\bar{f}\gamma_{\mu}\gamma_{5}f
O_{U}^{\mu}
\end{eqnarray}

\section{Cross sections for Moller scattering}

\begin{figure}
\includegraphics{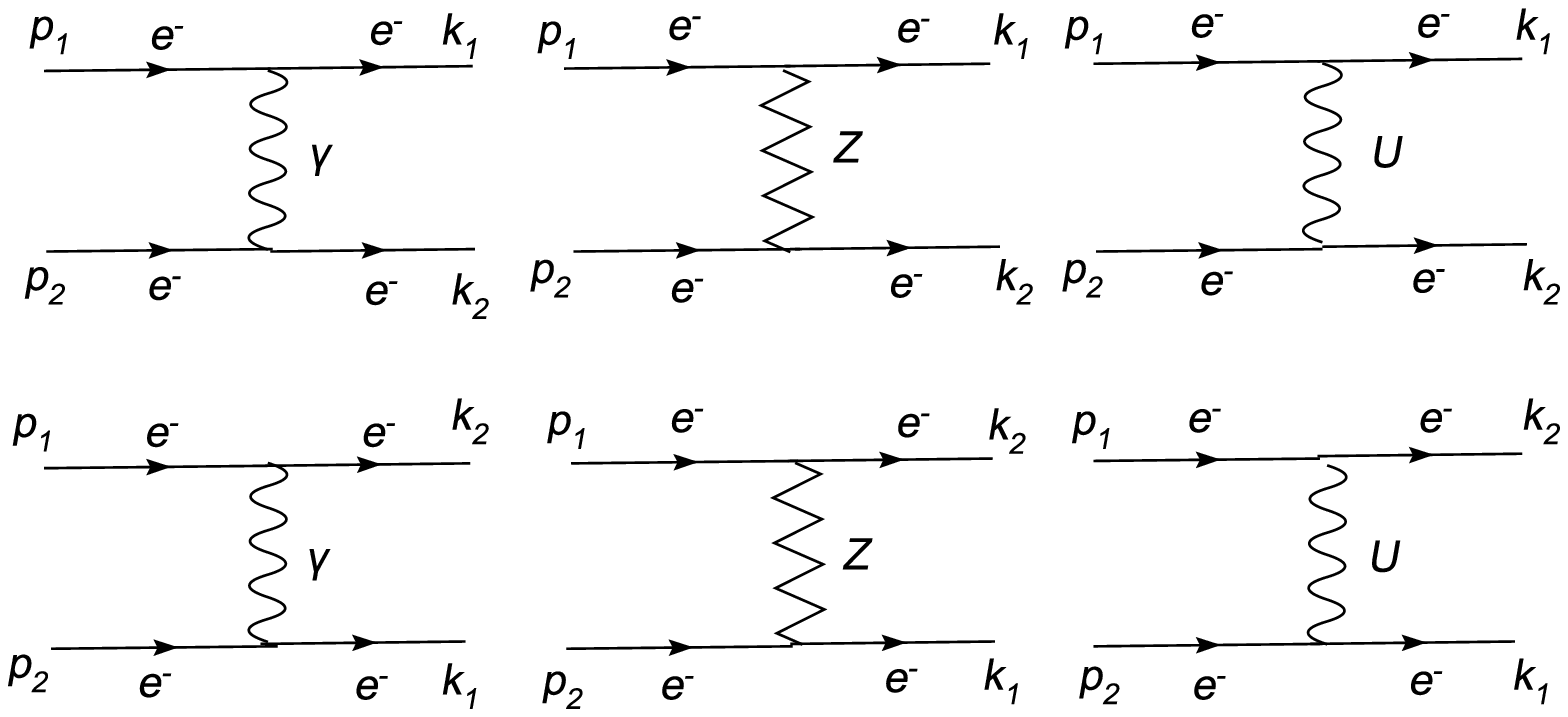}
\caption{ Tree-level Feynman diagrams for Moller scattering in the
presence of the couplings (5).\label{fig1}}
\end{figure}

In the presence of the couplings (5), Moller scattering is described
by the six t and u-channel tree-level diagrams in Fig.\ref{fig1}.
Two of them contain unparticle exchange and modify the SM
amplitudes.

The polarization summed scattering amplitude for Fig.\ref{fig1} is
given by,

\begin{eqnarray}
|M|^{2}=&&g_{e}^{4}A_{1}+\frac{g_{z}^{4}}{16}A_{2}+\frac{g_{e}^{2}g_{z}^{2}}{4}A_{3}
-c_{un}^{2}(-t)^{d_{U}-2}(-u)^{d_{U}-2}A_{4}+g_{e}^{2}c_{un}\left[
(-t)^{d_{U}-2}A_{5} \right. \nonumber \\&& \left.
 +(-u)^{d_{U}-2}A_{6}\right]+ \frac {g_{z}^{2}}{4}c_{un}\left[(-t)^{d_{U}-2}A_{7}
 +(-u)^{d_{U}-2}A_{8}\right]+c_{un}^{2}(-t)^{2d_{U}-4}A_{9}
 \nonumber \\
 &&+c_{un}^{2}(-u)^{2d_{U}-4}A_{10}
\end{eqnarray}
where,

\begin{eqnarray}
A_{1}=&&\frac{2(s^{4}+4ts^{3}+5t^{2}s^{2}+2t^{3}s+t^{4})}{t^{2}(s+t)^{2}}\\
\nonumber\\
A_{2}=&&\frac{1}{(m_{z}^{2}-t)^{2}(m_{z}^{2}-u)^{2}}\left\{\left[
(c_{A}^{2}-c_{V}^{2})^{2}\left(2t^{4}+4st^{3}+2m_{z}^{4}+6sm_{z}^{2}+2stm_{z}^{4}
\right. \right. \right. \nonumber \\&& \left. \left. \left.
+6s^{2}tm_{z}^{2}+8s^{3}t+s^{2}m_{z}^{4}+2s^{3}m_{z}^{2}+2s^{4}\right)+(5c_{A}^{4}s^{2}t^{2}
+6c_{A}^{2}c_{V}^{2}s^{2}t^{2}+5c_{V}^{4}s^{2}t^{2}) \right]
\right\}\\
\nonumber\\
A_{3}=&&\left[2\left(c_{A}^{2}(m_{z}^{2}s(-s^{2}+3ts+3t^{2})-2tu(2s^{2}+ts+t^{2}))
-c_{V}^{2}(s^{2}+ts+t^{2})(3sm_{z}^{2} \right.\right. \nonumber
\\&& \left. \left.
 +2(s^{2}+ts+t^{2}))\right)\right]/((m_{z}^{2}-t)t(s+t)(m_{z}^{2}-u))\\\nonumber\\
 A_{4}=&&-2(\lambda_{A}^{4}+6\lambda_{A}^{2}\lambda_{V}^{2}+\lambda_{V}^{4})s^{2}
\end{eqnarray}

\begin{eqnarray}
A_{5}=&&-\frac{2\left[\lambda_{V}^{2}(2s+t)(s^{2}+ts+t^{2})-\lambda_{A}^{2}t(3s^{2}+3ts+t^{2})
\right]}{tu}\\\nonumber\\
A_{6}=&&-\frac{2\left[(\lambda_{A}^{2}+\lambda_{V}^{2})s^{3}+(\lambda_{A}^{2}-\lambda_{V}^{2})
t^{3}\right]}{tu}\\\nonumber\\
A_{7}=&&\frac{1}{(m_{z}^{2}-t)(m_{z}^{2}-u)}\left\{2\left[-(c_{A}^{2}-c_{V}^{2})
(\lambda_{A}^{2}-\lambda_{V}^{2})\left(t^{3}+(m_{z}^{2}+3s)t^{2}+st(2m_{z}^{2}+3s)
\right) \right. \right. \nonumber\\&& \left. \left.
+s^{2}\left(-2s(c_{A}\lambda_{A}-c_{V}\lambda_{V})^{2}-((3\lambda_{A}^{2}
+\lambda_{V}^{2})c_{A}^{2}-8c_{V}c_{A}\lambda_{V}\lambda_{A}
\right.\right. \right. \nonumber\\&& \left. \left. \left.
+c_{V}^{2}m_{z}^{2}
(\lambda_{A}^{2}+3\lambda_{V}^{2}))\right)\right]\right\}\\\nonumber\\
A_{8}=&&\frac{1}{(m_{z}^{2}-t)(m_{z}^{2}-u)}\left\{2\left[(c_{A}^{2}-c_{V}^{2})
(\lambda_{A}^{2}-\lambda_{V}^{2})(t^{3}-m_{z}^{2}t^{2})-\left((c_{A}^{2}+c_{V}^{2})
(\lambda_{A}^{2}+\lambda_{V}^{2}) \right. \right. \right. \nonumber
\\&& \left. \left. \left. -4c_{V}c_{A}\lambda_{V}\lambda_{A}\right)s^{2}
(2m_{z}^{2}+s)\right]\right\}\\\nonumber\\
A_{9}=&&(\lambda_{A}^{2}-\lambda_{V}^{2})^{2}(t^{2}+2st)+2s^{2}
(\lambda_{A}^{2}+\lambda_{V}^{2})^{2}\\\nonumber\\
A_{10}=&&s^{2}(\lambda_{A}^{4}+6\lambda_{A}^{2}\lambda_{V}^{2}+\lambda_{V}^{4})
+t^{2}(\lambda_{A}^{2}-\lambda_{V}^{2})^{2}
\end{eqnarray}
\begin{eqnarray}
c_{un}=\frac{A_{d_{U}}}{2\sin{(\pi
d_{U})}\Lambda_{U}^{2d_{U}-2}},\,\,\,\,\,
c_{V}=-\frac{1}{2}+2\sin^{2}\theta_{W}, \,\,\,\,\,c_{A}=\frac{1}{2}
 ,\,\,\,\,\,g_{z}=\frac{g_{e}}{\sin\theta_{W}\cos\theta_{W}}
\end{eqnarray}

The Mandelstam parameters s, t and u are defined by,
$s=(p_{1}+p_{2})^{2}$, $t=(p_{1}-k_{1})^{2}$ and
$u=(p_{1}-k_{2})^{2}$. In the cross section calculations we impose a
cut $|cos\theta|<0.9$ on the scattering angle of one of the final
electrons in the C.M. frame.

\begin{figure}
\includegraphics{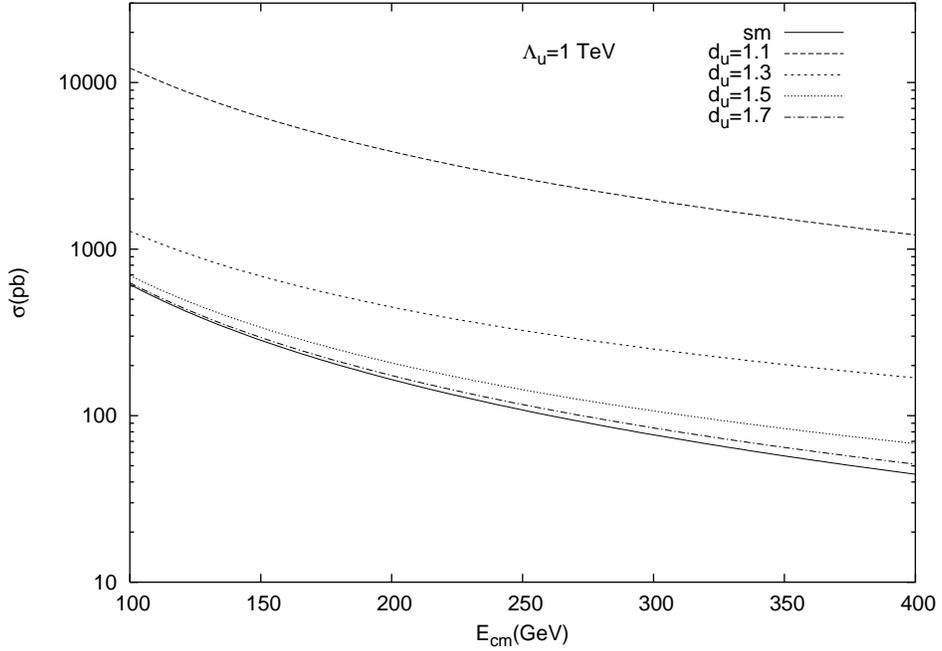}
\caption{ The total cross section for $\lambda_{V}=1$ and
$\lambda_{A}=0$ as a function of center of mass energy. The legends
are for different values of the scale dimension $d_{U}$.
$\Lambda_{U}$= 1 TeV. \label{fig2}}
\end{figure}

\begin{figure}
\includegraphics{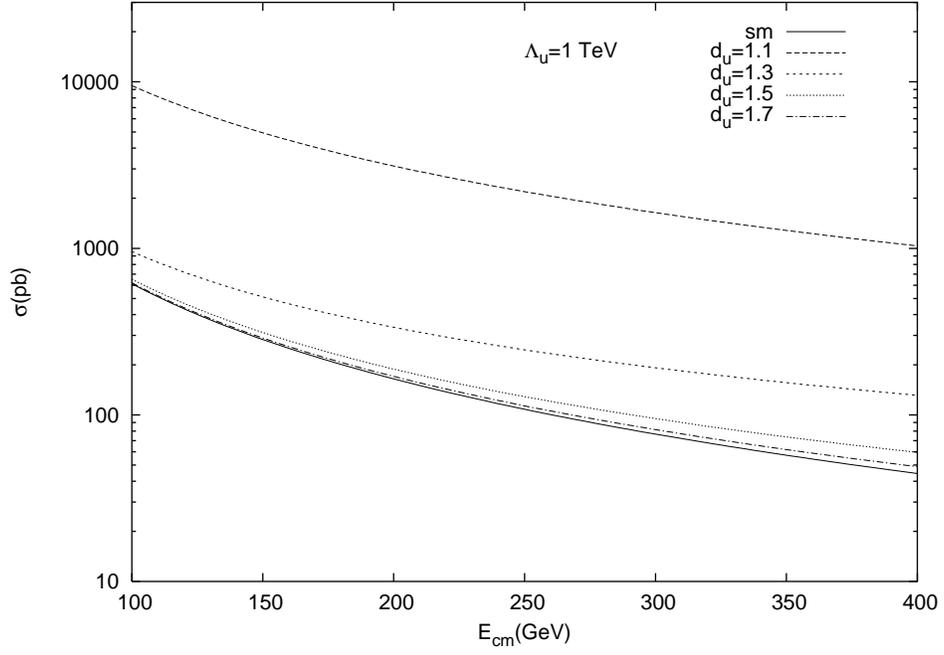}
\caption{ The same as Fig.2 but for $\lambda_{V}=0$ and
$\lambda_{A}=1$.\label{fig3}}
\end{figure}

\begin{figure}
\includegraphics{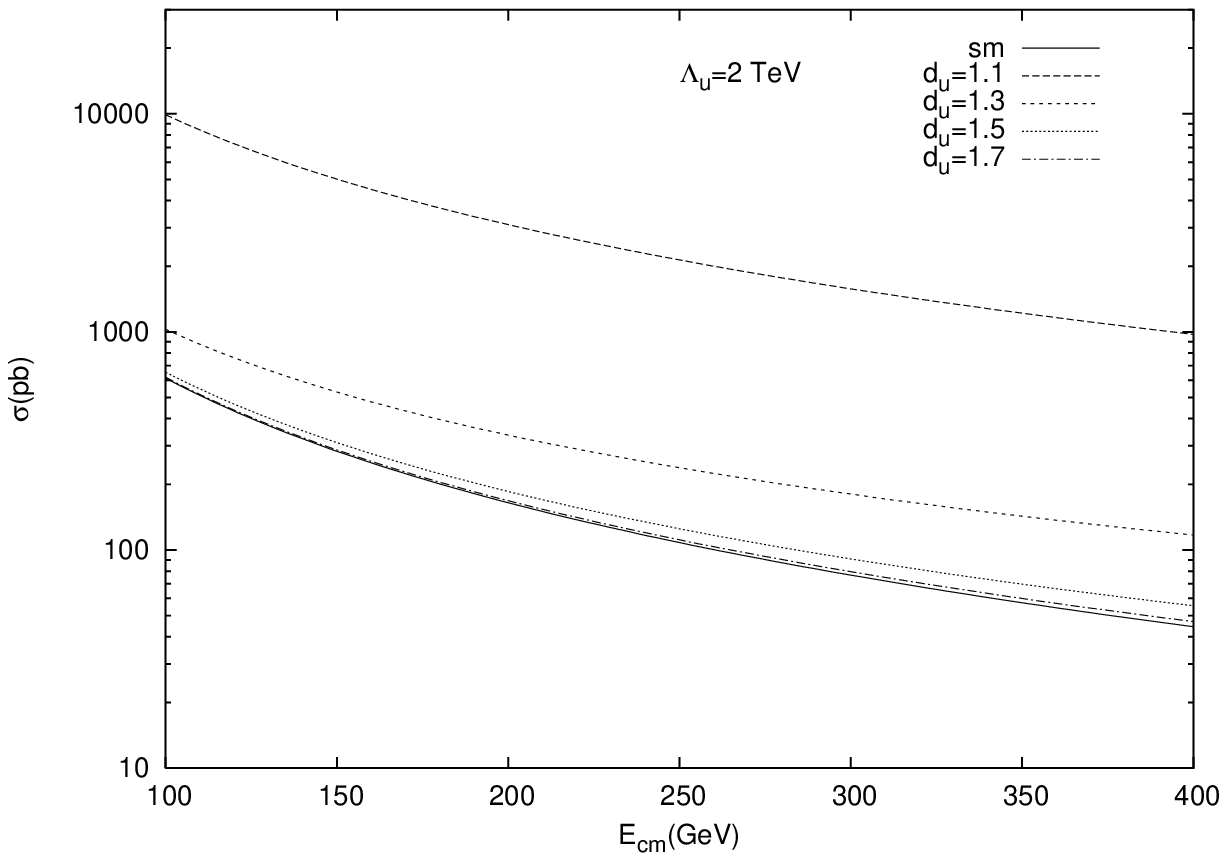}
\caption{ The same as Fig.2 but for $\Lambda_{U}$= 2
TeV.\label{fig4}}
\end{figure}

\begin{figure}
\includegraphics{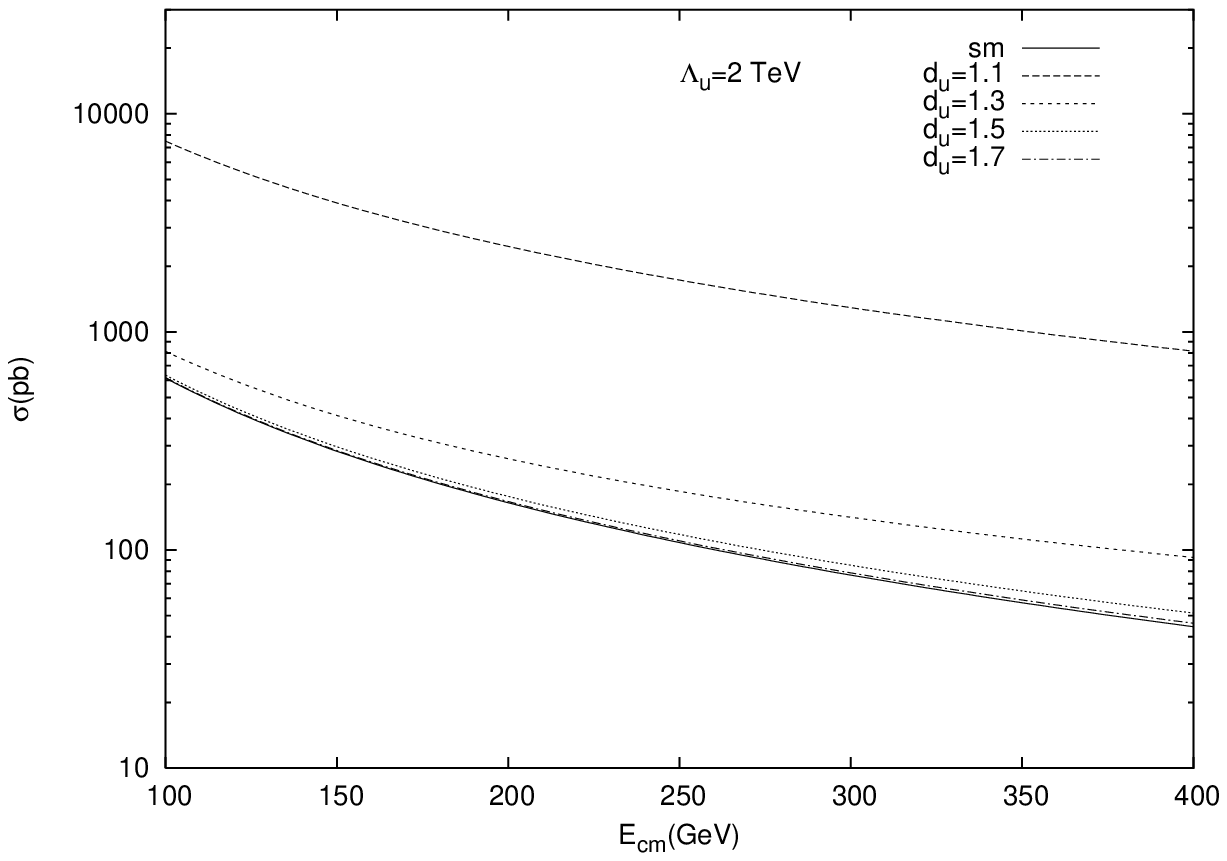}
\caption{ The same as Fig.3 but for $\Lambda_{U}$= 2
TeV.\label{fig5}}
\end{figure}

The behavior of the total cross section as a function of the center
of mass energy of the $e^{-}e^{-}$ system for $d_{U}=1.1, 1.3, 1.5,
1.7$ is shown in Fig.\ref{fig2}-\ref{fig5}. In the figures we
investigate the influence of the scale dimension $d_{U}$ on the
deviations of the total cross sections from their SM value for
$\Lambda_{U}=1$ TeV and  $\Lambda_{U}=2$ TeV. We omit a plot of the
cross section for $d_{U}=1.9$ since it is very close to the SM. One
can see from these figures that the deviations of the cross sections
grow as the energy increases.

In Fig.\ref{fig2} and Fig.\ref{fig4} we investigate the sensitivity
of the cross section to the vector coupling $\lambda_{V}$. So we set
$\lambda_{V}=1$ and $\lambda_{A}=0$. Similarly in Fig.\ref{fig3} and
Fig.\ref{fig5} we investigate the axial vector coupling
$\lambda_{A}$. We see from these figures that the cross section is
more sensitive to $\lambda_{V}$ than $\lambda_{A}$. For instance, in
Fig\ref{fig3} the cross section for $d_{U}=1.3$ at $E_{cm}$=400 GeV
increases by a factor of 3.0 when we compare with its SM value. On
the other hand in Fig.\ref{fig2} this increment is a factor of 3.8
for the same scale dimension $d_{U}=1.3$. We also see from
Fig.\ref{fig2}-\ref{fig5} that the deviation of the cross section
from its SM value increases with decreasing $d_{U}$. This is
reasonable since the $d_{U}$ dependent coefficient $c_{un}$ is
inversely proportional to the $(2d_{U}-2)$th power of the energy
scale $\Lambda_{U}$ (17). Therefore the contribution that comes from
the unparticle couplings drastically grow as the $d_{U}$ decreases.
To be precise for $\Lambda_{U}=\,1000\, GeV$,
$\frac{1}{\Lambda_{U}^{2d_{U}-2}}$ grows with a factor of 4000 as
$d_{U}$ decreases from 1.7 to 1.1.

\section{Constraints on the unparticle couplings}

A more detailed investigation of the unparticle couplings
$\lambda_{V}$ and $\lambda_{A}$ requires statistical analysis. To
this purpose we have obtained $95\%$ C.L. limits on $\lambda_{V}$
and $\lambda_{A}$ using a simple $\chi^{2}$ analysis at
$\sqrt{s}$=100, 300, 500 GeV and integrated luminosity $L_{int}$=50
and 500 $fb^{-1}$ without systematic errors. The $\chi^{2}$ function
is given by,

\begin{eqnarray}
\chi^{2}=\left(\frac{\sigma_{SM}-\sigma(\lambda_{V},\lambda_{A})}{\sigma_{SM}
\,\, \delta}\right)^{2}
\end{eqnarray}

where $\delta=\frac{1}{\sqrt{N}}$ is the statistical error. N is the
number of events. It is given by $N=L_{int}\sigma_{SM}$.

The limits for $\lambda_{V}$ and $\lambda_{A}$ are given in Tables
\ref{tab1}-\ref{tab4}. One can see from these tables that the lower
and upper bounds on the unparticle couplings are symmetric. The
decrease in $d_{U}$ highly improves the sensitivity limits. The most
sensitive results are obtained at $d_{U}=1.1$. This value of the
scale dimension improves the sensitivity limits of $\lambda_{V}$ by
a factor of 6 - 16 depending on the energy when we compare with
$d_{U}=1.7$ for $L_{int}$=50 $fb^{-1}$. This improvement is a factor
of 6 - 14 for the limits of $\lambda_{A}$, depending on the energy.

The energy dependence of the sensitivity limits are interesting. As
we have discussed in the previous section the deviation of the cross
sections grow as the energy increases. On the other hand the SM
cross section and therefore the number of events decreases with the
energy. Therefore it is very difficult to predict the behavior of
the limits without an explicit calculation. Explicit results are
given in the tables.

We see from the tables that the limits for the parameter
$\lambda_{V}$ are more sensitive than $\lambda_{A}$. For instance,
the sensitivity limit of $\lambda_{V}$ for $d_{U}=1.1$ is 1.7 times
restricted compared to $\lambda_{A}$ at $L_{int}$=50 $fb^{-1}$. This
factor is 1.75 at $L_{int}$=500 $fb^{-1}$.


\pagebreak

\begin{table}
\caption{Sensitivity of Moller scattering to $\lambda_{V}$ coupling
at 95\% C.L. for $L_{int}=50$ $fb^{-1}$ and $\Lambda_{U}=1000
 GeV.$ \label{tab1}}
\begin{ruledtabular}
\begin{tabular}{ccccc}
$\sqrt{s}(GeV)$&$d_{u}=1.1$&$d_{u}=1.3$&$d_{u}=1.5$&$d_{u}=1.7$ \\
\hline
 100&-0.007, 0.007 &-0.019,0.019&-0.046,0.046&-0.115,0.115\\
 300&-0.011,0.011 &-0.026,0.026&-0.052,0.052&-0.098,0.098\\
 500&-0.014,0.014 &-0.026,0.026&-0.048,0.048&-0.085,0.085\\
\end{tabular}
\end{ruledtabular}
\end{table}

\begin{table}
\caption{Sensitivity of  Moller scattering to $\lambda_{A}$ coupling
at 95\% C.L. for $L_{int}=50$ $fb^{-1}$ and $\Lambda_{U}=1000
 GeV.$ \label{tab2}}
\begin{ruledtabular}
\begin{tabular}{ccccc}
$\sqrt{s}(GeV)$&$d_{u}=1.1$&$d_{u}=1.3$&$d_{u}=1.5$&$d_{u}=1.7$ \\
\hline
100&-0.012,0.012& -0.030,0.030&-0.076,0.076&-0.168,0.168\\
300&-0.016,0.016&-0.035,0.035&-0.071,0.071&-0.130,0.130\\
500&-0.018,0.018&-0.034,0.034&-0.063,0.063&-0.104,0.104\\
\end{tabular}
\end{ruledtabular}
\end{table}

\begin{table}
\caption{Sensitivity of Moller scattering to $\lambda_{V}$ coupling
at 95\% C.L. for $L_{int}=500$ $fb^{-1}$ and $\Lambda_{U}=1000
 GeV.$ \label{tab3}}
\begin{ruledtabular}
\begin{tabular}{ccccc}
$\sqrt{s}(GeV)$&$d_{u}=1.1$&$d_{u}=1.3$&$d_{u}=1.5$&$d_{u}=1.7$ \\
\hline
100&-0.004,0.004&-0.012,0.012&-0.032,0.032&-0.080,0.080\\
300&-0.006,0.006&-0.014,0.014&-0.032,0.032&-0.062,0.062\\
500&-0.008,0.008&-0.016,0.016&-0.029,0.029&-0.047,0.047\\

\end{tabular}
\end{ruledtabular}
\end{table}

\begin{table}
\caption{Sensitivity of Moller scattering to $\lambda_{A}$ coupling
at 95\% C.L. for $L_{int}=500$ $fb^{-1}$ and $\Lambda_{U}=1000
 GeV.$ \label{tab4}}
\begin{ruledtabular}
\begin{tabular}{ccccc}
$\sqrt{s}(GeV)$&$d_{u}=1.1$&$d_{u}=1.3$&$d_{u}=1.5$&$d_{u}=1.7$ \\
\hline
100&-0.007,0.007&-0.017,0.017&-0.041,0.041&-0.091,0.091\\
300&-0.010,0.010&-0.022,0.022&-0.042,0.042&-0.078,0.078\\
500&-0.010,0.010&-0.019,0.019&-0.035,0.035&-0.055,0.055\\

\end{tabular}
\end{ruledtabular}
\end{table}

\end{document}